# ANALYSIS OF THE RELEVANCE OF THE FILTERED RADIATIVE TRANSFER EQUATION TERMS FOR LARGE EDDY SIMULATION OF TURBULENCE-RADIATION INTERACTION

Maxime Roger*, Pedro J. Coelho*§ and Carlos B. da Silva*
*Mechanical Engineering Department, Instituto Superior Técnico
Technical University of Lisbon, Av. Rovisco Pais, 1049-001 Lisboa, Portugal
§Correspondence author. Fax: +351 218475545 Email: pedro.coelho@ist.utl.pt

**ABSTRACT** An analysis of the turbulence-radiation interaction in the framework of large eddy simulation (LES) is presented. Direct numerical simulation (DNS) of statistical steady forced homogeneous isotropic turbulence is used to evaluate the relevance of the unclosed terms of the filtered radiative transfer equation (RTE) and to study the influence of parameters like the optical thickness and the turbulence intensity. LES without subgrid-scale models of the filtered RTE has also been investigated. Neglecting the subgrid-scale fluctuations of radiation has proved to be an accurate assumption in various cases, especially for the absorption terms of radiation. However, turbulence-radiation interaction effects increase significantly with the optical thickness based on the size of the filter or with the turbulence intensity, and consequently subgrid-scale modelling should be developed in cases where the optical thickness is not thin and the turbulence intensity higher than 20%.

## NOMENCLATURE

$F$ Flatness
$k_{max}$ Maximum resolved turbulent wave number
$L$ Size of the radiation domain
$L_{11}$ Turbulent integral length scale
$N_b$ Number of narrow bands
$N_Q$ Number of quadrature points
$Re_\lambda$ Reynolds number based on Taylor micro-scale
$s$ Coordinate along the optical path
$S$ Skewness
Sc Schmidt number
$X$ Molar fraction

*Greek Symbols*
$\delta$ Size of the control volume in the DNS grid
$\Delta$ Size of the box filter
$\Delta\nu$ Narrow band width
$\eta$ Kolmogorov micro-scale
$\kappa$ Absorption coefficient
$\nu$ Molecular viscosity
$\tau_\Delta$ Optical thickness based on the size of the filter

*Subscripts*
$b$ Blackbody
$\nu$ Wavenumber

*Notations*
$\langle Q \rangle$ Time-average of $Q$
$\langle Q'^2 \rangle$ Variance of $Q$
$\overline{Q}$ Spatially filtered quantity
$\overline{Q''}$ Filtered residual quantity

## INTRODUCTION

In the numerical simulation of reactive flows, radiative transfer often plays a relevant role, and turbulence-radiation interaction (TRI) must be taken into account in most cases [*e.g.,* Coelho 2007]. The TRI has generally been investigated using Reynolds-averaged Navier-Stokes methods [Coelho 2004], probability density functions approach [Mazumder and Modest 1999, Liu *et al.* 2004] and recently Direct Numerical Simulation (DNS) [Wu *et al.* 2005, Deshmukh *et al.* 2007]. Although DNS is a powerful tool to provide fundamental insight on turbulent flows [Pope 2000], it is too computational expensive for practical applications, and is restricted to moderate Reynolds numbers and/or simple geometries. In this context, and with the increase of computer power, large eddy simulation (LES) is becoming an attractive approach to achieve high accuracy at an affordable computational cost for solving moderately complex problems of practical usefulness in the Computational Fluid Dynamics and combustion communities [Janicka and Sadiki 2005]. To our knowledge, almost no work has been done so far to simulate the TRI in LES of reactive flows. Thermal radiation has been coupled to LES of combustion system in [Desjardin and Frankel 1999, Haworth *et al.* 2005, Jones and Paul 2005] but without subgrid-scale models for the resolution of the filtered radiative transfer equation (RTE). Poitou *et al.* [2007] have made an a priori study from DNS of a reactive shear layer to evaluate the TRI from the emission part of radiation and have tested models based on Taylor development to reconstruct the correlation between the temperature and the absorption coefficient in a flame.

In this context, the present study reports an analysis of the relevance of the unclosed terms of the filtered RTE that may help the future development of closure models to include the TRI in LES of reactive flows. Our approach has been based on DNS calculation of statistical steady forced homogeneous isotropic turbulence, which offers the simplest possible turbulent flow configuration, and is consequently a useful tool for a parametric analysis and for developing and assessing new turbulence models. An a-priori analysis from DNS of homogeneous isotropic turbulence for LES calculation of the TRI has been carried out and parameters like the optical thickness, the filter size and the turbulence intensities are investigated. Finally the filtered RTE has been solved without subgrid-scale models, and some results will be presented.

## THEORY

**The Filtered Radiative Transfer Equation** The RTE in the case of an emitting-absorbing and non-scattering medium may be written as:

$$\frac{dI_\nu}{ds} = -\kappa_\nu I_\nu + \kappa_\nu I_{b\nu} \tag{1}$$

where $\nu$ is the wavenumber, $I_\nu$ the spectral radiation intensity, $s$ the coordinate along the optical path, $\kappa_\nu$ the spectral absorption coefficient and $I_{b\nu}$ the Planck function. In LES, the larger turbulent motions are directly represented whereas the effect of the smaller scale motions are modelled [Pope 2000]. A low-pass filtering operation is performed so that the resulting filtered quantity can be adequately resolved on a coarser grid. In this study, we have considered the box filter function defined such that $\overline{Q}$ represents the quantity $Q$ averaged over a cell of volume $\Delta^3$ (where $\Delta$ is the filter size):

$$\overline{Q} = \frac{1}{\Delta^3}\iiint_{\Delta^3} Q(\vec{x},t)\, d\vec{x} \tag{2}$$

By applying the filtering operation to equation (1), we obtain the filtered RTE:

$$\frac{d\bar{I}_\nu}{d\,s} = -\overline{\kappa_\nu\, I_\nu} + \overline{\kappa_\nu\, I_{b\nu}} \tag{3}$$

After expanding $\kappa_\nu$, $I_\nu$ and $I_{b\nu}$ on the right hand side of equation (3) as the sum of a filtered quantity and a fluctuating component, this equation may be expressed as

$$\frac{d\bar{I}_\nu}{d\,s} = -\overline{\bar{\kappa}_\nu\, \bar{I}_\nu} - \left(\overline{\kappa''_\nu\, \bar{I}_\nu} + \overline{\bar{\kappa}_\nu\, I''_\nu} + \overline{\kappa''_\nu\, I''_\nu}\right) + \overline{\bar{\kappa}_\nu\, \bar{I}_{b\nu}} + \left(\overline{\kappa''_\nu\, \bar{I}_{b\nu}} + \overline{\bar{\kappa}_\nu\, I''_{b\nu}} + \overline{\kappa''_\nu\, I''_{b\nu}}\right) \tag{4}$$

where $I''_\nu$, $I''_{b\nu}$ and $\kappa''_\nu$ are the residual fields. The six terms in parentheses must be modelled in order to close the equation and to include TRI effects in LES of turbulent flows. In this study, we have focused on radiative quantities integrated over the spectrum like the Planck mean absorption coefficient $\kappa_P$ and the incident mean absorption coefficient $\kappa_G$ which are defined as follows

$$\kappa_P = \frac{\int_0^{+\infty} \kappa_\nu\, I_{b\nu}\, d\nu}{\int_0^{+\infty} I_{b\nu}\, d\nu} \tag{5}$$

$$\kappa_G = \frac{\int_0^{+\infty} \kappa_\nu\, G_\nu\, d\nu}{\int_0^{+\infty} G_\nu\, d\nu} = \frac{\int_0^{+\infty} \kappa_\nu\, I_\nu\, d\nu}{\int_0^{+\infty} I_\nu\, d\nu} \tag{6}$$

where $G_\nu$ is the spectral incident radiation. The second equality in equation (6) only holds for isotropic radiation, as in the present analysis. In this case, the equation for the total filtered intensity may then be written as:

$$\frac{d\bar{I}}{ds} = -\overline{\kappa_G\, I} + \overline{\kappa_P\, I_b} = -\overline{\bar{\kappa}_G\, \bar{I}} - \left(\overline{\kappa''_G\, \bar{I}} + \overline{\bar{\kappa}_G\, I''} + \overline{\kappa''_G\, I''}\right) + \overline{\bar{\kappa}_P\, \bar{I}_b} + \left(\overline{\kappa''_P\, \bar{I}_b} + \overline{\bar{\kappa}_P\, I''_b} + \overline{\kappa''_P\, I''_b}\right) \tag{7}$$

The assumption of isotropic radiation simplifies the physical problem presented in the remainder of this study. In practical problems, this assumption is approximately valid for perfectly stirred reactors and for flows of burnt gases in a post-combustion zone. Furthermore, Eq. (7) is similar to Eq. (4), which does not require the assumption of isotropic radiation, and therefore it is expected that the conclusions drawn from the analysis of Eq. (7) remain valid for Eq. (4).

_The filtered divergence of the radiative flux_ The divergence of the radiative heat flux may be expressed as

$$\nabla\vec{q} = \int_0^{+\infty} \left[4\pi\, \kappa_\nu\, I_{b\nu} - \int_{4\pi} \kappa_\nu\, I_\nu\, d\Omega\right] d\nu \tag{8}$$

Assuming isotropic radiation, we can rewrite the divergence of the radiative heat flux in the following form

$$\nabla\vec{q} = 4\pi\left[\kappa_P\, I_b - \kappa_G\, I\right] \tag{9}$$

Applying the filtering operation to equation (9) yields

$$\nabla \overline{q} = 4\pi \left[\overline{\kappa_P I_b} - \overline{\kappa_G I}\right] = 4\pi \left[\overline{\kappa}_P \overline{I}_b + \left(\overline{\kappa''_P \overline{I}_b} + \overline{\overline{\kappa}_P I''_b} + \overline{\kappa''_P I''_b}\right) - \overline{\kappa}_G \overline{I} - \left(\overline{\kappa''_G \overline{I}} + \overline{\overline{\kappa}_G I''} + \overline{\kappa''_G I''}\right)\right] \quad (10)$$

In equations (7) or (10), the terms depending on $\kappa_P$ and $I_b$ will be referred as the emission terms and those depending on $\kappa_G$ and $I$ as the absorption terms of the filtered RTE.

**Large Eddy Simulation of the RTE without subgrid-scale models** The simplest way to close equation (4) is to suppose that there are no subgrid-scale fluctuations concerning radiation. In this assumption, and in the case of a medium composed by a mixture of $H_2O$ and $CO_2$, the filtered RTE may be written as

$$\frac{d\overline{I}_\nu}{ds} \approx -\overline{\kappa}_\nu \overline{I}_\nu + \overline{\kappa}_\nu \overline{I}_{b\nu} \approx -\kappa_\nu(\overline{T}, \overline{X}_{CO_2}, \overline{X}_{H_2O})\, \overline{I}_\nu + \kappa_\nu(\overline{T}, \overline{X}_{CO_2}, \overline{X}_{H_2O})\, I_{b\nu}(\overline{T}) \quad (11)$$

The six terms in parentheses in equation (4), *i.e.* the subgrid-scale fluctuations are neglected and it is further assumed that $\overline{\kappa}_\nu = \kappa_\nu(\overline{T}, \overline{X}_{CO_2}, \overline{X}_{H_2O})$ and $\overline{I}_{b\nu} = I_{b\nu}(\overline{T})$.

## COMPUTATIONAL DETAILS

**Direct Numerical Simulation of Homogeneous Isotropic Turbulence.** The DNS are carried out using a pseudo-spectral code in which the temporal advancement is made with an explicit 3rd order Runge-Kutta scheme. A passive scalar field transport equation with Schmidt number equal to 0.7 is also solved along with the Navier-Stokes equations. The physical domain is a cubic box of size $2\pi$ and the simulations use $192^3$ collocation points. The large scales of both the velocity and scalar fields were forced in order to sustain the turbulence. Table 1 lists some of the flow parameters, which characterize the behaviour of the small scales of motion. The skewness (S) and flatness (F) factors of the velocity derivative are similar to the ones found in high Reynolds number turbulence. Moreover, the energy and scalar variance spectra (not shown) exhibit an inertial range with about one decade, and furthermore the Reynolds number based on the Taylor micro-scale is close to 100. These results show that at least the inertial and dissipative of scales of motion from the present simulation are characteristic of the flow field in mixing layers, jets, and wakes, at the far field and at moderate Reynolds number. More details on this DNS data bank can be found in [da Silva and Pereira 2007] and also in [da Silva *et al.* 2006]. In the following, the size of the control volumes in the DNS grid is denoted by $\delta$.

Table 1
Parameters of the turbulent flow

| $Re_\lambda$ | $\nu$ | Sc | $k_{max}\eta$ | $L_{11}$ | $\eta$ | $S$ | $F$ |
|---|---|---|---|---|---|---|---|
| 95.6 | 0.006 | 0.7 | 1.8 | 1.24 | 0.028 | -0.49 | 4.63 |

*Rescaling the DNS data into the radiation domain.* The data obtained from homogeneous isotropic turbulence simulations in the cubic box are rescaled into the radiation domain by assuming kinematics similarity between the two flows [*e.g.* da Silva *et al.* 2006]. This assumption yields the following expression of the instantaneous temperature field $T$ used in the radiative heat transfer calculations as a function of the DNS temperature field $T_{DNS}$

$$T(\vec{x}) = \langle T_{rad} \rangle + T_{DNS}(\vec{x}) \sqrt{\frac{\langle T'^2_{rad} \rangle}{\langle T'^2_{DNS} \rangle}} \tag{12}$$

Here, $\langle T_{rad} \rangle$ is the time-averaged temperature prescribed at point $\vec{x}$, and $\langle T'^2_{rad} \rangle$ and $\langle T'^2_{DNS} \rangle$ are the variance of the temperature fields prescribed for the radiation calculations and computed from DNS, respectively. This rescaling from DNS data to radiation calculations should be manipulated carefully, especially in cases where the fluctuation intensities are large. Sometimes, the temperature is beyond the physical range. In such a case, we may assume that $T(\vec{x}_i) = T_{\max}$ in points where the temperature exceeds the maximum possible temperature of the system, $T_{\max}$ (and the same for points $\vec{x}_j$ where $T(\vec{x}_j)$ is lower than $T_{\min}$). This assumption remains acceptable as long as the statistics of the temperature field are not significantly modified, which is the case if the turbulence intensity, defined by $\sqrt{\langle T'^2_{rad} \rangle} / \langle T_{rad} \rangle$, does not exceed 30%.

**Radiative Transfer Calculation** The integral form of the RTE may be written as [Modest 2003]

$$I_\nu(s) = I_\nu(0) \exp\left( - \int_0^s \kappa_\nu(s')\, ds' \right) + \int_0^s \kappa_\nu(s')\, I_{b\nu}(s') \exp\left( - \int_{s''}^s \kappa_\nu(s'')\, ds'' \right) ds' \tag{13}$$

Periodic boundary conditions were employed in the present work to be consistent with the boundary conditions used in DNS. The periodicity was enforced by setting $I_\nu(L) = I_\nu(0)$ where $L$ is the length of the radiation domain. Equation (13) has been discretized by dividing the optical paths into elements and interpolating the temperature and chemical composition from the DNS data using cubic splines. The integrals in equation (13) are numerically evaluated using Simpson's rule. The integration over the spectrum has been done using the correlated-k distribution method [Goody *et al.* 1989] (CK approximation). In this method, the spectrum is divided into narrow bands such that $\int_0^{+\infty} I_\nu\, d\nu \approx \sum_{i=1}^{N_b} I_{\Delta\nu_i}$, and the absorption coefficient is reordered within every band into a smooth monotonically increasing function. The averaged radiation intensity over a band takes the following form

$$I_{\Delta\nu} = \frac{1}{\Delta\nu} \int_{\Delta\nu} I_\nu\, d\nu = \int_0^\infty I(k)\, f(k)\, dk = \int_0^1 I(g)\, dg \tag{14}$$

where $f(k)$ is the probability density function of the absorption coefficient in the considered band and $g(k)$ is the cumulative k-distribution function. The integral in equation (14) is evaluated using Gaussian quadrature, which yields the following relation (for a mixture of two absorbing species)

$$I_{\Delta\nu} = \sum_{i=1}^{N_b} \sum_{j=1}^{N_Q} \sum_{k=1}^{N_Q} \omega_j\, \omega_k\, I_{\Delta\nu_i,j,k}\, \Delta\nu_i \tag{15}$$

where $\omega_j$ and $\omega_k$ are quadrature weights, $N_Q$ is the number of quadrature points and $I_{\Delta\nu_i,j,k}$ is the radiative intensity for the $i$th band and for quadrature points $j$ and $k$. The parameters needed for the CK approximation are taken from the data of Soufiani and Taine [1997]. It was assumed that the temperature and the absorbing species are fully correlated, which is consistent with the laminar flamelet combustion model.

## RESULTS AND DISCUSSION

Although radiation from $H_2O$ could easily be included in the calculations, only radiation from carbon dioxide is considered here in order to reduce computational requirements. This is sufficient for our main goal in this paper, namely to investigate the TRI in LES. The standard radiative transfer calculations were carried out assuming that the mean temperature of the medium is 1500 K and the mean molar fraction of $CO_2$ is 0.5. The root mean square of temperature and carbon dioxide molar fraction are 300 K and 0.1 respectively (the turbulence intensity is equal to 20%). The length $L$ of the radiation domain is defined in order to satisfy the prescribed optical thickness of the medium τ (estimated by $\tau = \langle \kappa_P L \rangle$) which is equal to 10. In the following calculations, these standard values remain unchanged unless other values are indicated.

**Filtered RTE terms evaluation** Figure 1 presents the temperature profile along an arbitrarily chosen line of sight, parallel to an axis of the cubic domain, and composed by 192 nodes of the DNS grid. Figures 2 and 3 show an example of the estimation of the various terms of the filtered RTE along the same line of sight. A strong correlation between the radiation and the temperature profiles is observed. The ratio $\Delta / \delta$ (where $\delta$ is the distance between neighbouring nodes of the DNS grid and $\Delta$ the size of the filter) is equal to the number of grid nodes within the box filter domain. The optical thickness based on the size of the filter $\tau_\Delta$ is then given by the relation $\tau_\Delta = (\tau / 192) \times (\Delta / \delta)$. Figure 2 shows the estimation of the emission terms with a filter of size $\Delta = 4\delta$, which corresponds to an optical thickness based on the size of the filter equal to 0.21. We observe that, as a first approximation, the subgrid-scale correlations can be neglected, the three terms $\overline{\kappa_P'' \bar{I}_b}$, $\overline{\bar{\kappa}_P I_b''}$ and $\overline{\kappa_P'' I_b''}$ being distinctly smaller than $\overline{\bar{\kappa}_P \bar{I}_b}$. The same observation can be made for the absorption terms in figure 3. However, this approximation should be used carefully, because the four terms $\overline{\kappa_P'' \bar{I}_b}$, $\overline{\bar{\kappa}_P I_b''}$, $\overline{\kappa_G'' \bar{I}}$ and $\overline{\bar{\kappa}_G I''}$ may increase locally and consequently neglecting the subgrid-scale fluctuations may not be accurate anymore in cases where high precision is needed (especially in the case of the emission terms).

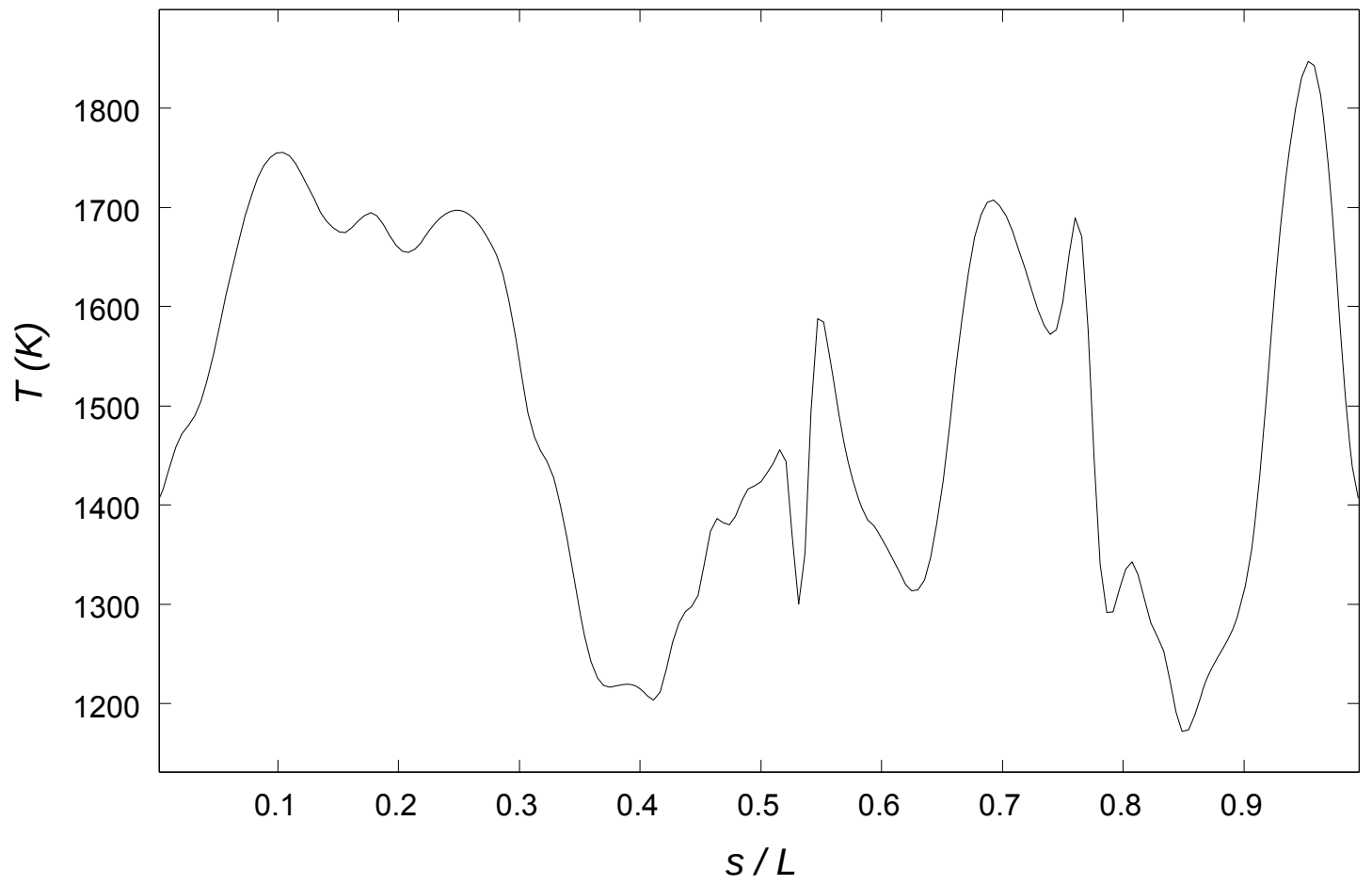

Figure 1. Temperature profile along a line of sight in the cubic domain.

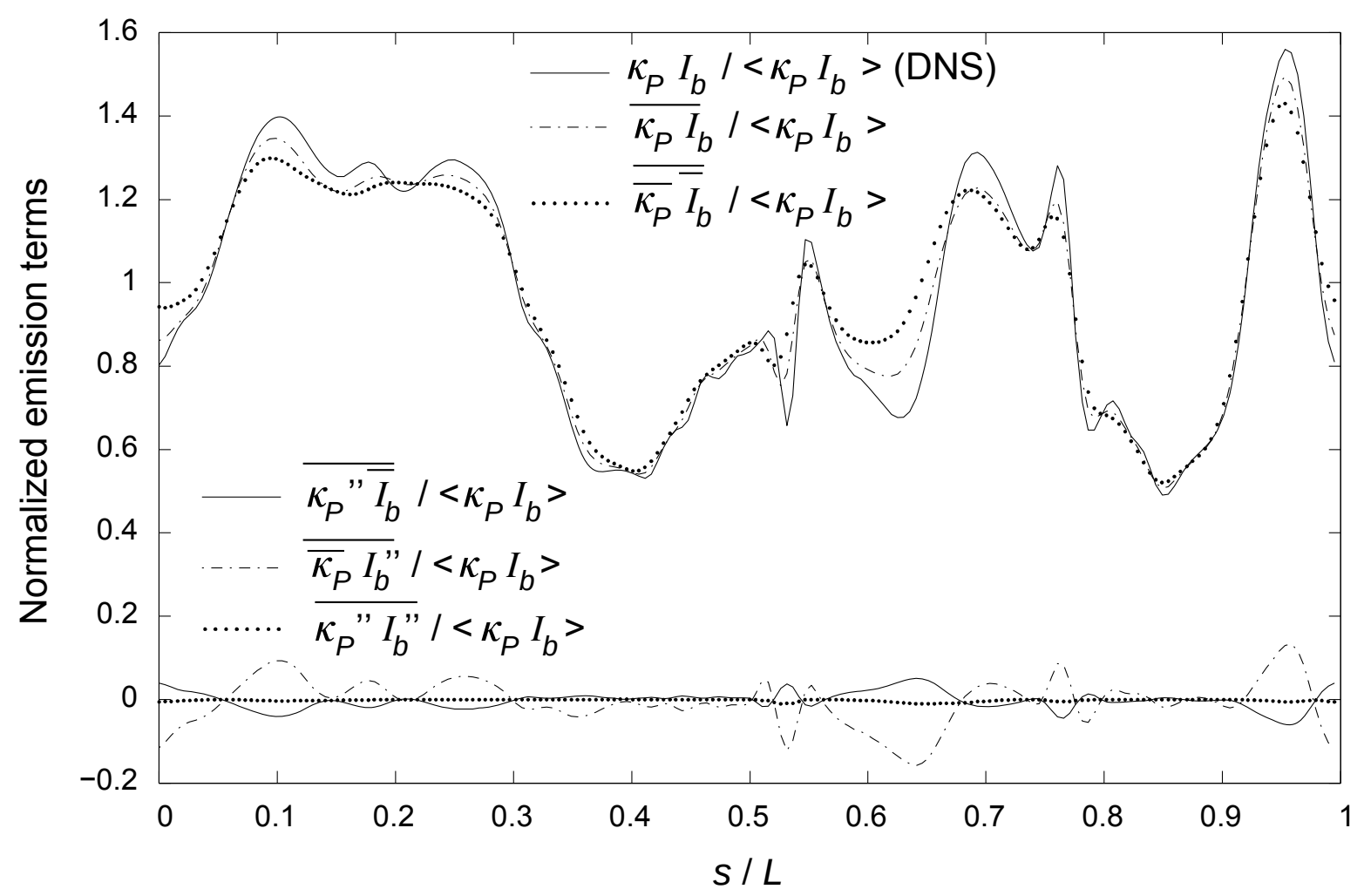

Figure 2. Normalized filtered RTE emission terms along a line of sight with a box filter of size $\Delta = 4\delta$ .

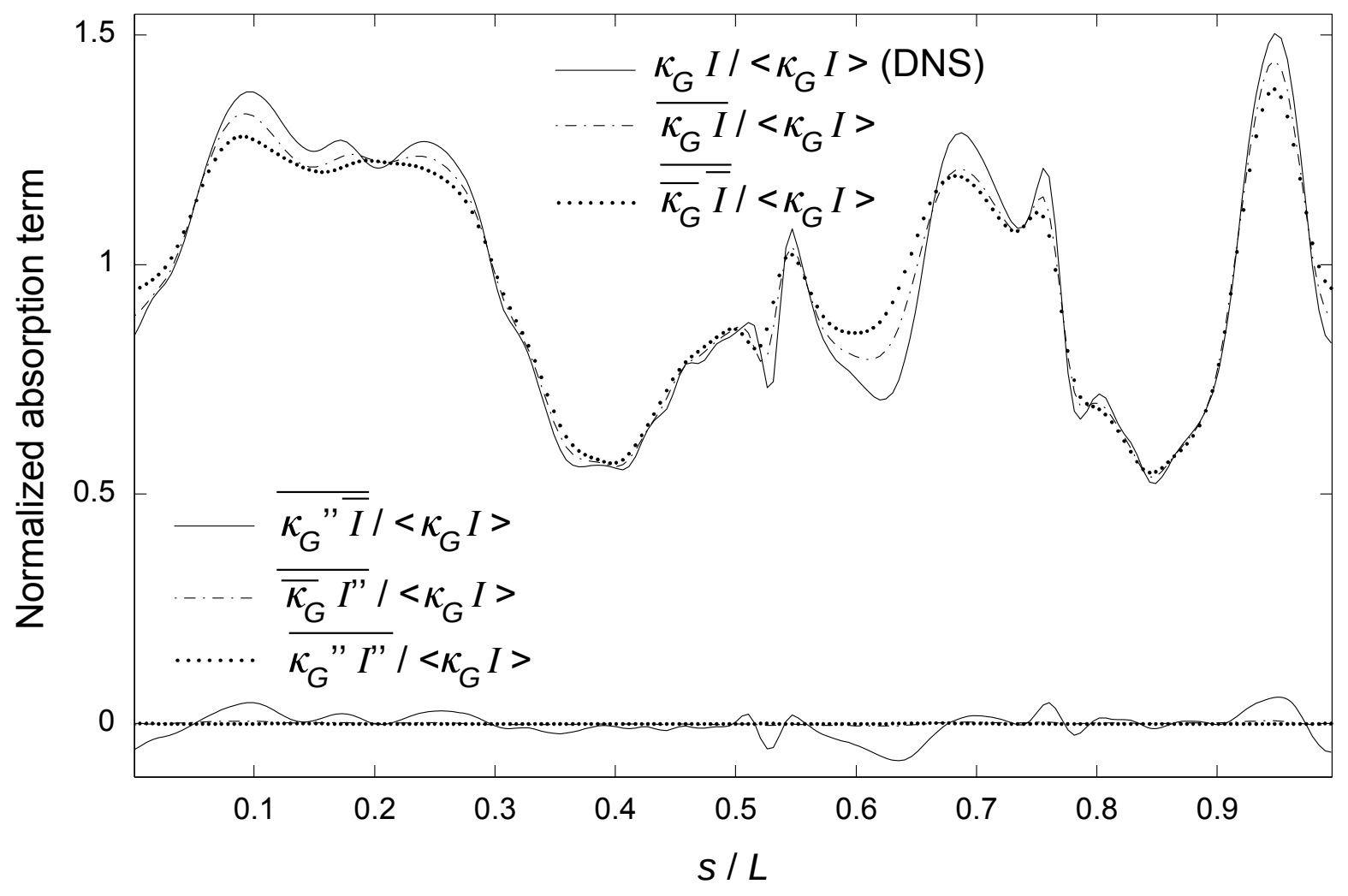

Figure 3. Normalized filtered RTE absorption terms along a line of sight with a box filter of size $\Delta = 4\delta$ .

<u>*Influence of filter size*</u> In Figures 4 and 5, the various terms of the filtered RTE have been estimated along the same line of sight but with a box filter equal to $\Delta$=32$\delta$ (optical thickness $\tau_\Delta$= 1.67). Obviously, the subgrid-scale fluctuations are more important in this case, and the assumption consisting in neglecting the subgrid-scale correlation between the absorption coefficient and the radiation intensity is no longer valid. Comparing Figures 4 and 5, it is concluded that the subgrid-scale fluctuations are more important for the emission terms than for the absorption terms.

These observations are confirmed by the results reported in Tables 2 and 3 where the mean values of the emission (Table 2) and absorption (Table 3) terms are presented for various filter sizes. The mean values were estimated over the $192^3$ points of the DNS grid[§]. Table 2 shows that for filters with an optical thickness $\tau_\Delta$ lower than 1 (filters of size $\Delta$ / $\delta$ equal to 2, 4 and 8), the error that results from neglecting the fluctuations is lower than 5% on average. In the case of filters with an optical thickness of the order of 1 or higher, the mean value $\left\langle \overline{\overline{\kappa_P}\, \overline{I}_b} / \overline{\kappa_P\, I_b} \right\rangle$ is significantly different from 1, and consequently models must be developed to take into account the correlations between the absorption coefficient and the Planck function. In the case of the absorption terms, Table 3 shows that the need for subgrid-scale models is less important than for the emission terms, confirming the tendency shown in Figures 3 and 5.

[§] In statistical steady (forced) homogeneous isotropic turbulence, the time-averaged quantities are statistically equal to the spatial-averaged quantities, which allow the evaluation of the mean value of a quantity with one instantaneous field over all the points of the cubic domain.

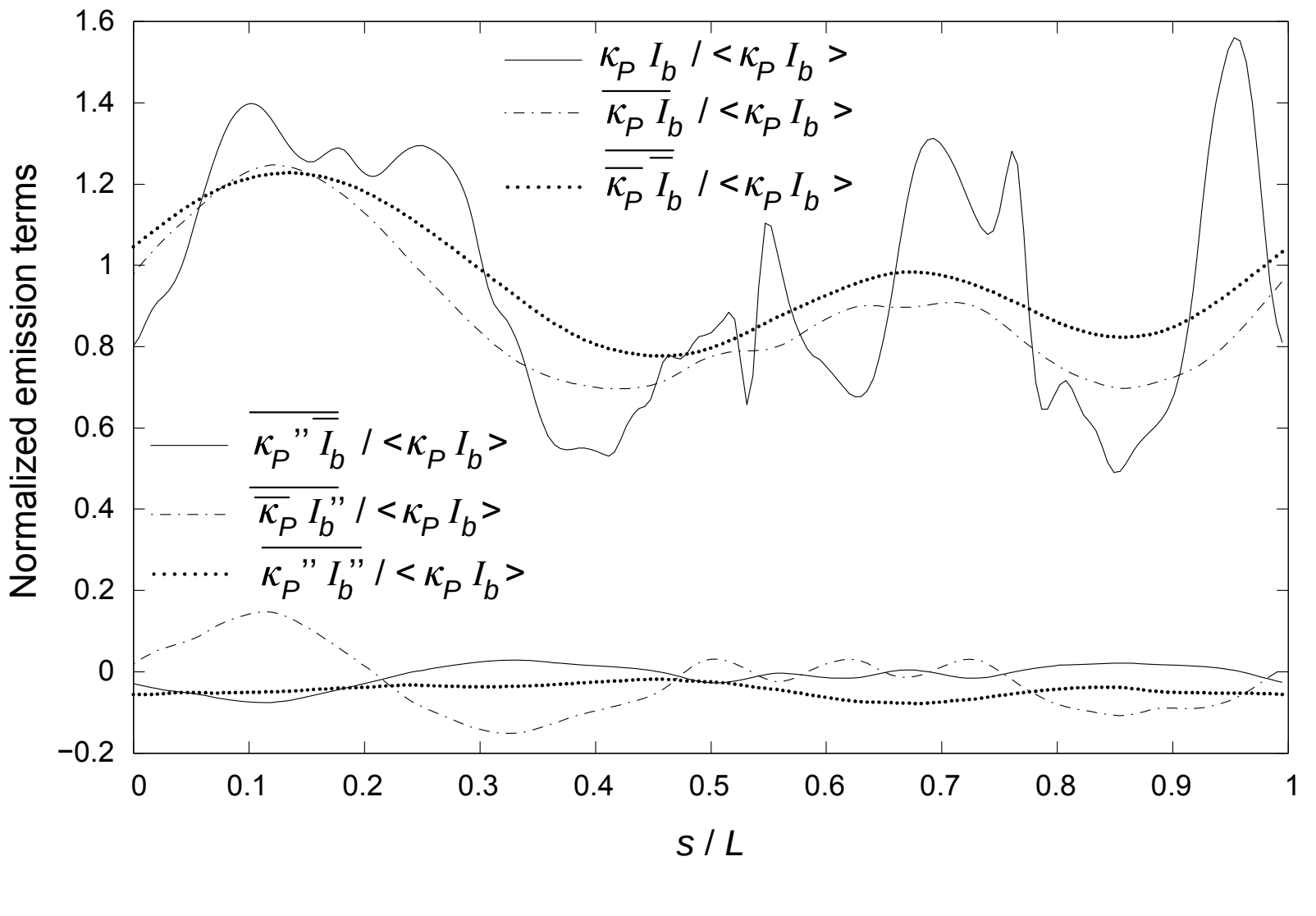

Figure 4. Normalized filtered RTE emission terms along a line of sight with a box filter of size $\Delta$=32 $\delta$.

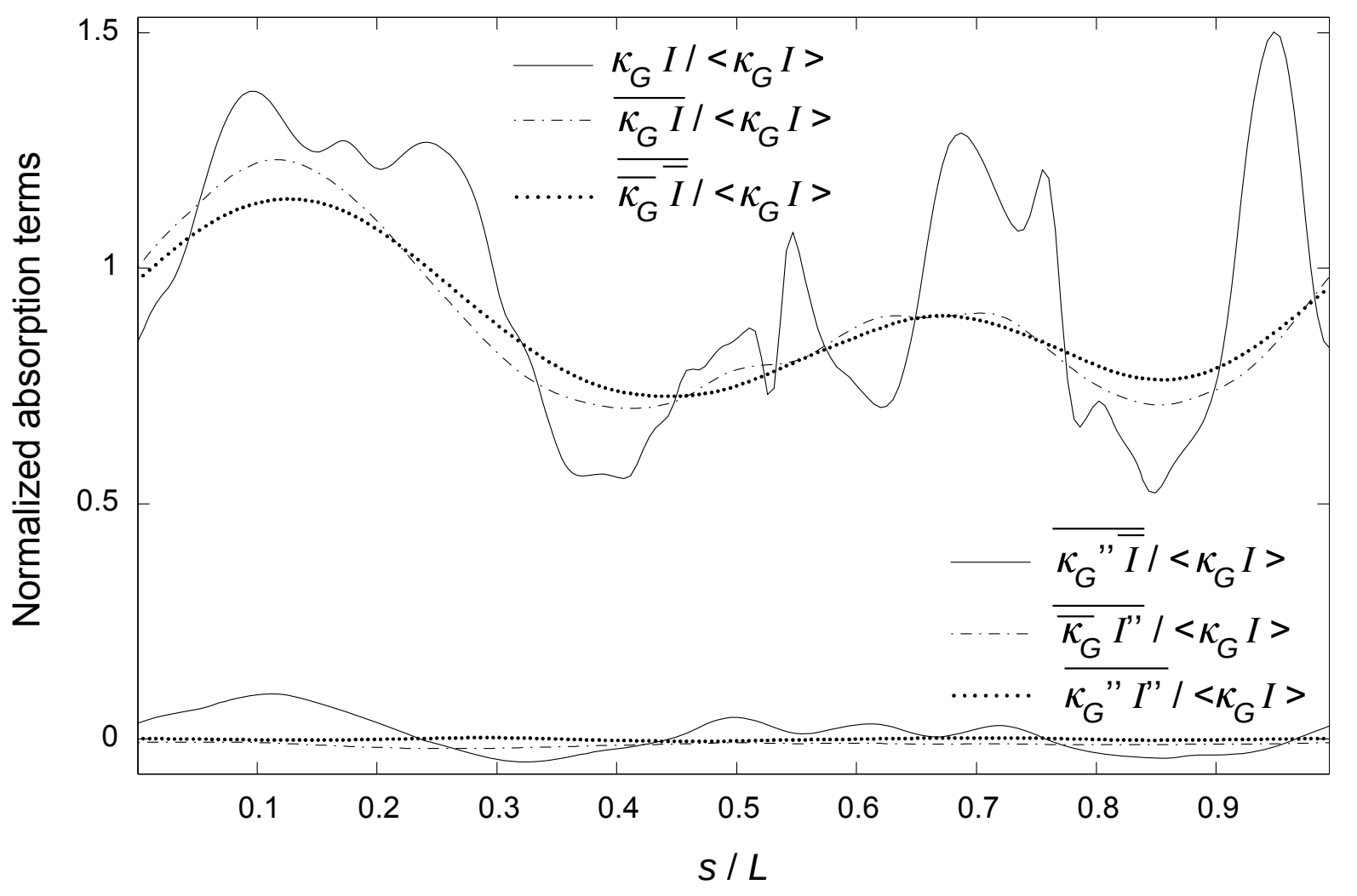

Figure 5. Normalized filtered RTE absorption terms along a line of sight with a box filter of size $\Delta$=32 $\delta$.

*Correlation between the absorption coefficient and the blackbody radiation intensity* In order to identify if locally important subgrid-scale correlations may appear in the system, Figures 6 and 7 show the joint-probability density functions between the total emission terms of the filtered RTE,

Table 2
Mean value of the normalized emission terms of the divergence of radiative flux

| $\Delta/\delta$<br>$\tau_\Delta$ | 2<br>0.104 | 4<br>0.208 | 8<br>0.417 | 16<br>0.833 | 32<br>1.67 |
|---|---|---|---|---|---|
| $\left\langle \frac{\overline{\overline{\kappa}_P \overline{I}_b}}{\overline{\kappa_P I_b}} \right\rangle$ | 1.00522 | 1.0169 | 1.0452 | 1.0929 | 1.156 |
| $\left\langle \frac{\overline{\kappa_P'' \overline{I}_b}}{\overline{\kappa_P I_b}} \right\rangle$ | 0.00126 | -0.000449 | -0.00482 | -0.0107 | -0.0162 |
| $\left\langle \frac{\overline{\overline{\kappa}_P I_b''}}{\overline{\kappa_P I_b}} \right\rangle$ | -0.00437 | -0.0138 | -0.0326 | -0.0528 | -0.0658 |
| $\left\langle \frac{\overline{\kappa_P'' I_b''}}{\overline{\kappa_P I_b}} \right\rangle$ | -0.00211 | -0.00259 | -0.00775 | -0.0294 | -0.0742 |

Table 3
Mean value of the normalized absorption terms of the divergence of radiative flux

| $\Delta/\delta$ | 2 | 4 | 8 | 16 | 32 |
|---|---|---|---|---|---|
| $\left\langle \frac{\overline{\overline{\kappa}_G \overline{I}}}{\overline{\kappa_G I}} \right\rangle$ | 1.00258 | 1.00623 | 1.0131 | 1.0211 | 1.0299 |
| $\left\langle \frac{\overline{\kappa_G'' \overline{I}}}{\overline{\kappa_G I}} \right\rangle$ | -0.00225 | -0.00537 | -0.0107 | -0.0147 | -0.0159 |
| $\left\langle \frac{\overline{\overline{\kappa}_G I''}}{\overline{\kappa_G I}} \right\rangle$ | -0.000275 | -0.000818 | -0.00239 | -0.00588 | -0.0108 |
| $\left\langle \frac{\overline{\kappa_G'' I''}}{\overline{\kappa_G I}} \right\rangle$ | $-4.95\times10^{-5}$ | $-4.40\times10^{-5}$ | $-4.78\times10^{-5}$ | $-4.64\times10^{-4}$ | -0.00320 |

$\overline{\kappa_P I_b}$, and the emission terms if the subgrid-scale correlations are neglected, $\overline{\kappa}_P \overline{I}_b$. In Figure 6, where a box filter of size $\Delta=2\delta$ is used, the joint-probability density function is almost a straight line. The two quantities are very close to each other, and consequently the subgrid-scale fluctuations of the emission term may be neglected. In Figure 7, where a box filter of size $\Delta=32\delta$ is used, the joint-probability density function is obviously more spread than in Figure 6, which means that important differences between $\overline{\kappa_P I_b}$ and $\overline{\kappa}_P \overline{I}_b$ may occur, emphasizing the need of models for the emission terms that account for the subgrid-scale fluctuations. In the case of the absorption terms, the joint-probability density function (which is not presented here) shows the same evolution with an increase of the filter size but with a less pronounced spread. This confirms that subgrid-scale fluctuations are less important, which is explained by the fact that the incident mean absorption coefficient and the radiation intensity are not local quantities, *i.e.*, they depend on the temperature and species concentration along the optical path, and consequently they are less influenced by local turbulent fluctuations.

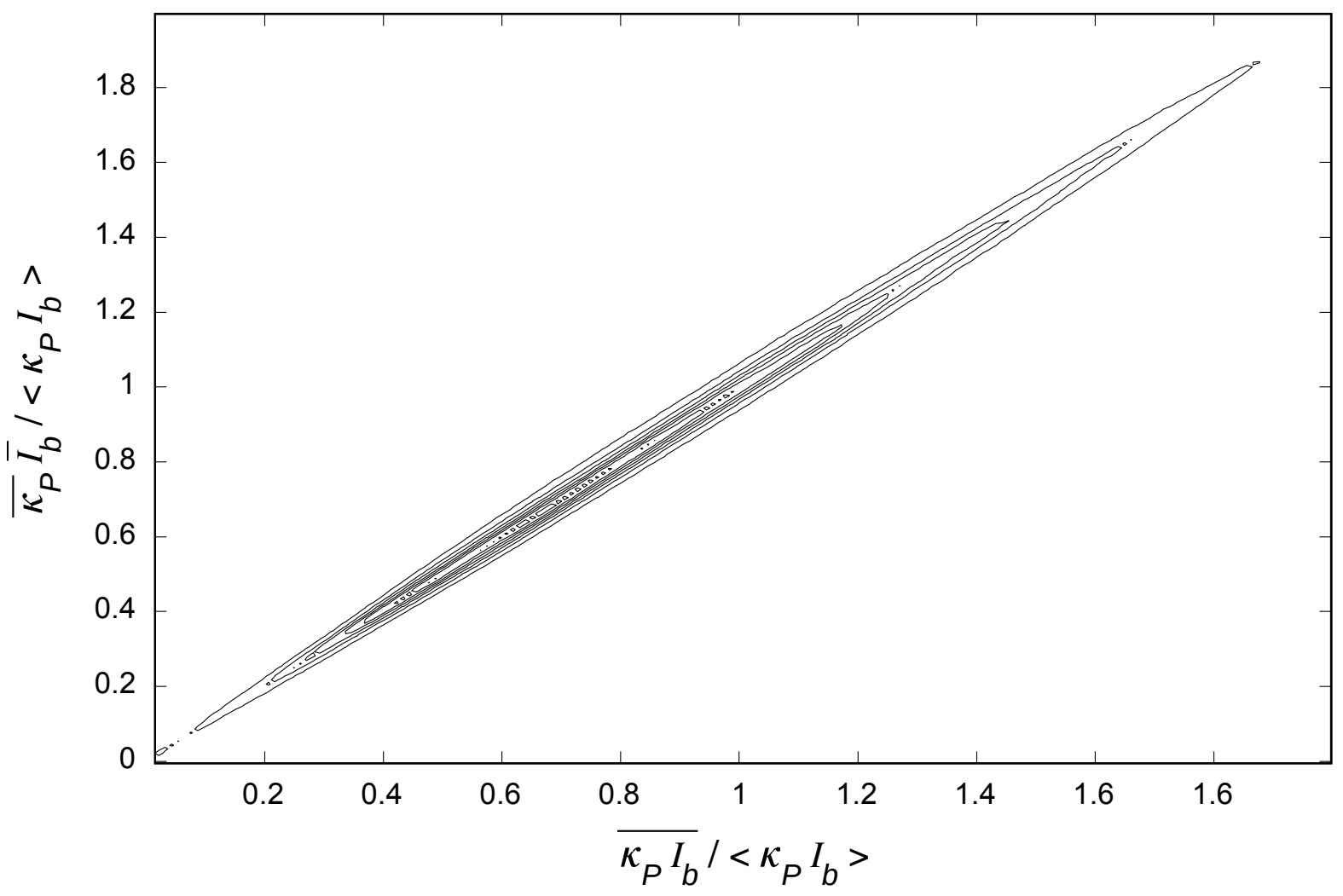

Figure 6. Joint $\overline{\kappa_P I_b}$ and $\bar{\kappa}_P \bar{I}_b$ probability density function (filter size Δ=2*δ*).

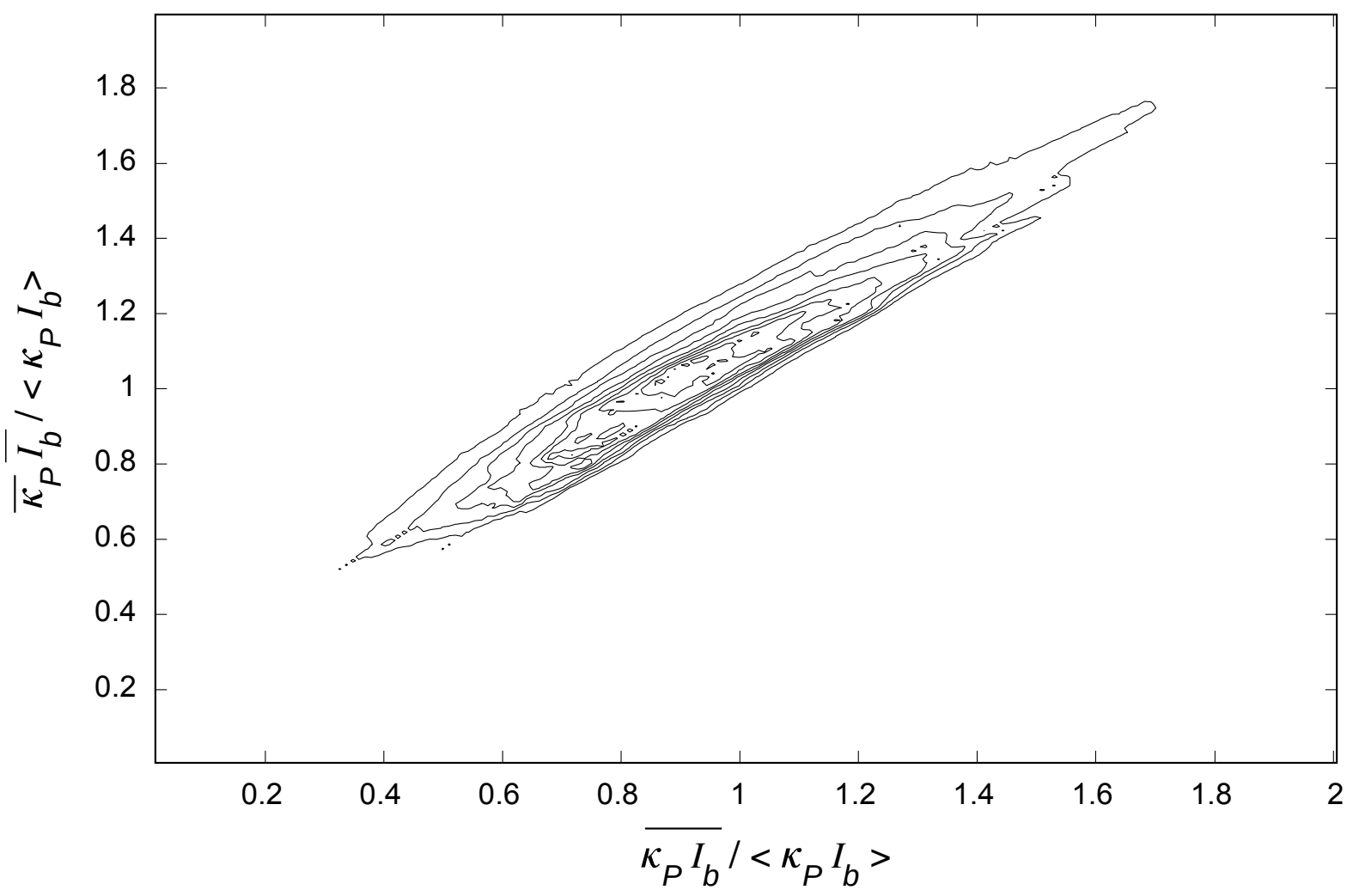

Figure 7. Joint $\overline{\kappa_P I_b}$ and $\bar{\kappa}_P \bar{I}_b$ probability density function (filter size Δ=32*δ*).

**Influence of the Turbulence Intensity** The filtered divergence of the radiative heat flux given by equation (10) may be rewritten as

$$\overline{\nabla \vec{q}} = 4\pi\left[\bar{\kappa}_P \bar{I}_b - \left(\bar{\kappa}_P \bar{I}_b - \overline{\kappa_P I_b}\right) - \bar{\kappa}_G \bar{I} + \left(\bar{\kappa}_G \bar{I} - \overline{\kappa_G I}\right)\right] \tag{16}$$

The terms in parentheses indicate the magnitude of the unresolved small scale fluctuations on the emission and absorption terms. So, we now introduce two quantities, $R_{\kappa_P I_b}$ and $R_{\kappa_G I}$, in order to evaluate how important the unresolved fluctuations of the radiation emission and absorption terms are:

$$R_{\kappa_P I_b} = \left\langle \frac{\bar{\kappa}_P \, \bar{I}_b - \overline{\kappa_P \, I_b}}{\overline{\kappa_P \, I_b}} \right\rangle \tag{17}$$

$$R_{\kappa_G I} = \left\langle \frac{\bar{\kappa}_G \, \bar{I} - \overline{\kappa_G \, I}}{\overline{\kappa_G \, I}} \right\rangle \tag{18}$$

In the remainder of this study, these quantities will be given in percentage.

Tables 4 and 5 show the results obtained for three turbulence intensities (10%, 20% and 30%). We observe an important influence on the subgrid-scale fluctuations, which increase with the turbulence intensity, for both the emission term (table 4) and for the absorption term (table 5) of the radiative heat flux. We also remark that a high turbulence intensity combined with a high optical thickness $\tau_\Delta$ (on the order of 1) yields a strong increase of subgrid-scale correlations which cannot be neglected anymore in these cases.

Table 4
Influence of the turbulence intensity in the subgrid-scale fluctuations of the emission term

| $\Delta / \delta$ | 2 | 4 | 8 | 16 | 32 |
|---|---|---|---|---|---|
| $R_{\kappa_P I_b}$ (10%) | 0.0547% | 0.246% | 0.780% | 1.84% | 3.39% |
| $R_{\kappa_P I_b}$ (20%) | 0.206% | 0.936% | 3.00% | 7.12% | 13.2% |
| $R_{\kappa_P I_b}$ (30%) | 0.312% | 1.52% | 5.28% | 13.2% | 25.0% |

Table 5
Influence of the turbulence intensity in the subgrid-scale fluctuations of the absorption term

| $\Delta / \delta$ | 2 | 4 | 8 | 16 | 32 |
|---|---|---|---|---|---|
| $R_{\kappa_G I}$ (10%) | 0.000968% | 0.00265% | 0.0208% | 0.0819% | 0.257% |
| $R_{\kappa_G I}$ (20%) | 0.00531% | 0.0265% | 0.109% | 0.387% | 1.07% |
| $R_{\kappa_G I}$ (30%) | 0.0130% | 0.0643% | 0.254% | 0.818% | 1.99% |

**LES calculation without subgrid-scale models** Figure 8 presents an example of LES estimation of the absorption terms without subgrid-scale models along a line of sight[§]. The filter domain is optically thin ($\tau_\Delta$=0.208, $\Delta$=4$\delta$). We observe that the two curves of $\overline{\kappa_G \, I}$ (filtered DNS results) and $\bar{\kappa}_G \, \bar{I}$ (estimated from LES) are almost merged, confirming that for filters with a thin optical thickness (lower than 1), the subgrid-scale correlations may be neglected. This observation is also valid for the emission term (which is not presented here). In figure 9, the filter has now an optical thickness of 0.833, *i.e.* Δ=16$\delta$. We observe that the two curves are very close to each other and the assumption consisting of neglecting subgrid-scale fluctuations is still accurate. However, we note that local differences between $\overline{\kappa_G I}$ and $\bar{\kappa}_G \bar{I}$ may appear, *e.g.* around $s/L$=0.6 in the line of sight presented, and so one should be careful. In figure 10, the filter has now an optical thickness of 1.67 (Δ=32$\delta$), and $\overline{\kappa_G I}$ and $\bar{\kappa}_G \bar{I}$ differ. Therefore, the subgrid-scale correlations between the absorption coefficient and the radiation intensity must be modelled (this observation is also valid for the emission term).

[§] The same line of sight in the cubic box is used for figures 1, 2, 3, 4, 5, 8, 9 and 10.

## CONCLUSIONS

An analysis of the relevance of turbulence-radiation interaction in the framework of large eddy simulation is presented. A parametric study based on direct numerical simulation of steady homogeneous isotropic turbulence is done in order to shed some light on a future modelling of the subgrid-scale fluctuations of radiative quantities. The influence of the filter size and its corresponding optical thickness, and the influence of the turbulence intensity are investigated. Finally, large eddy simulation without subgrid-scale models of the radiative transfer equation has also been carried out.

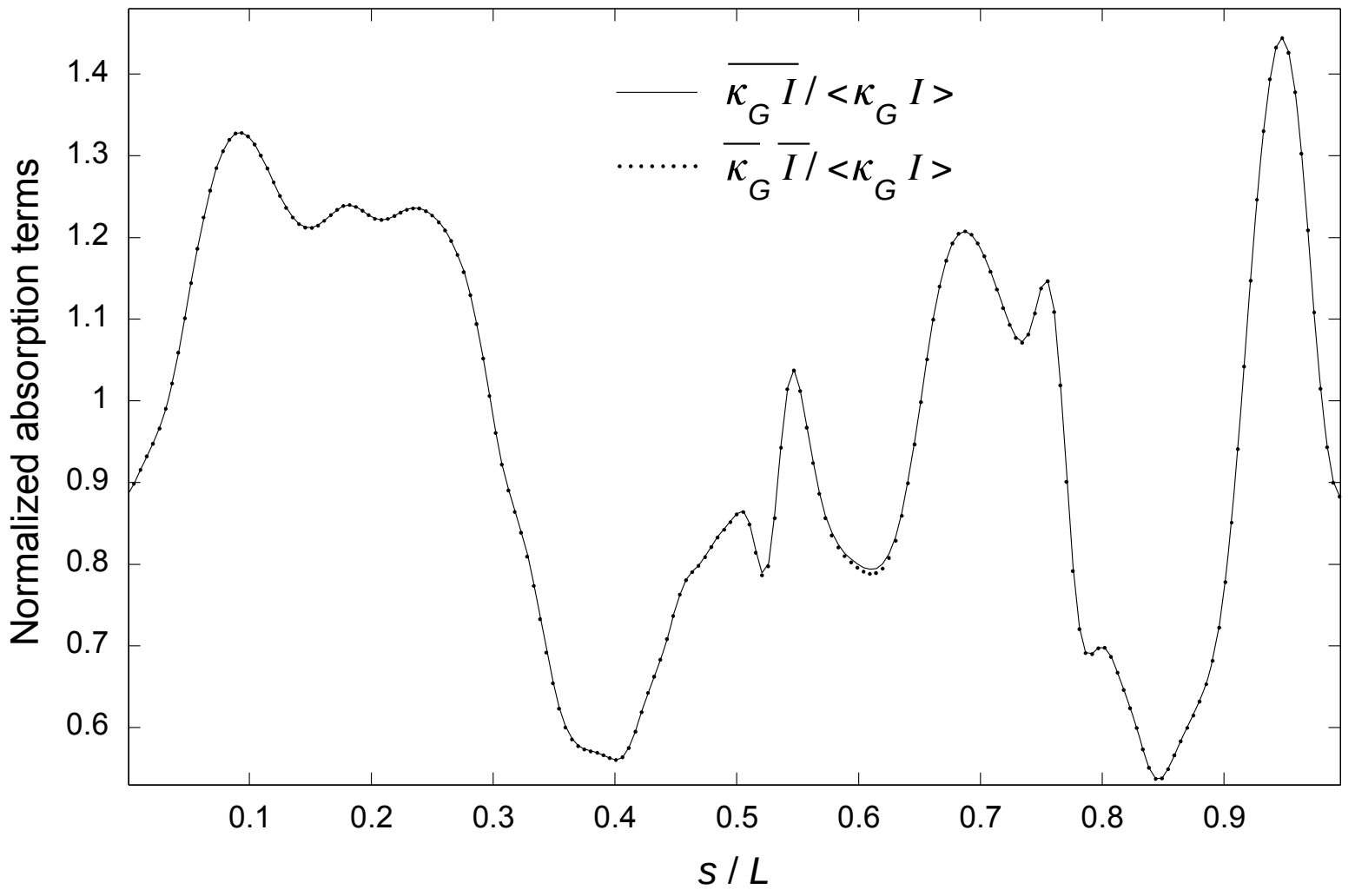

Figure 8. Absorption terms along a line of sight estimated by LES without subgrid-scale models compared with the filtered absorption terms determined from DNS (the filter size is Δ=4$\delta$).

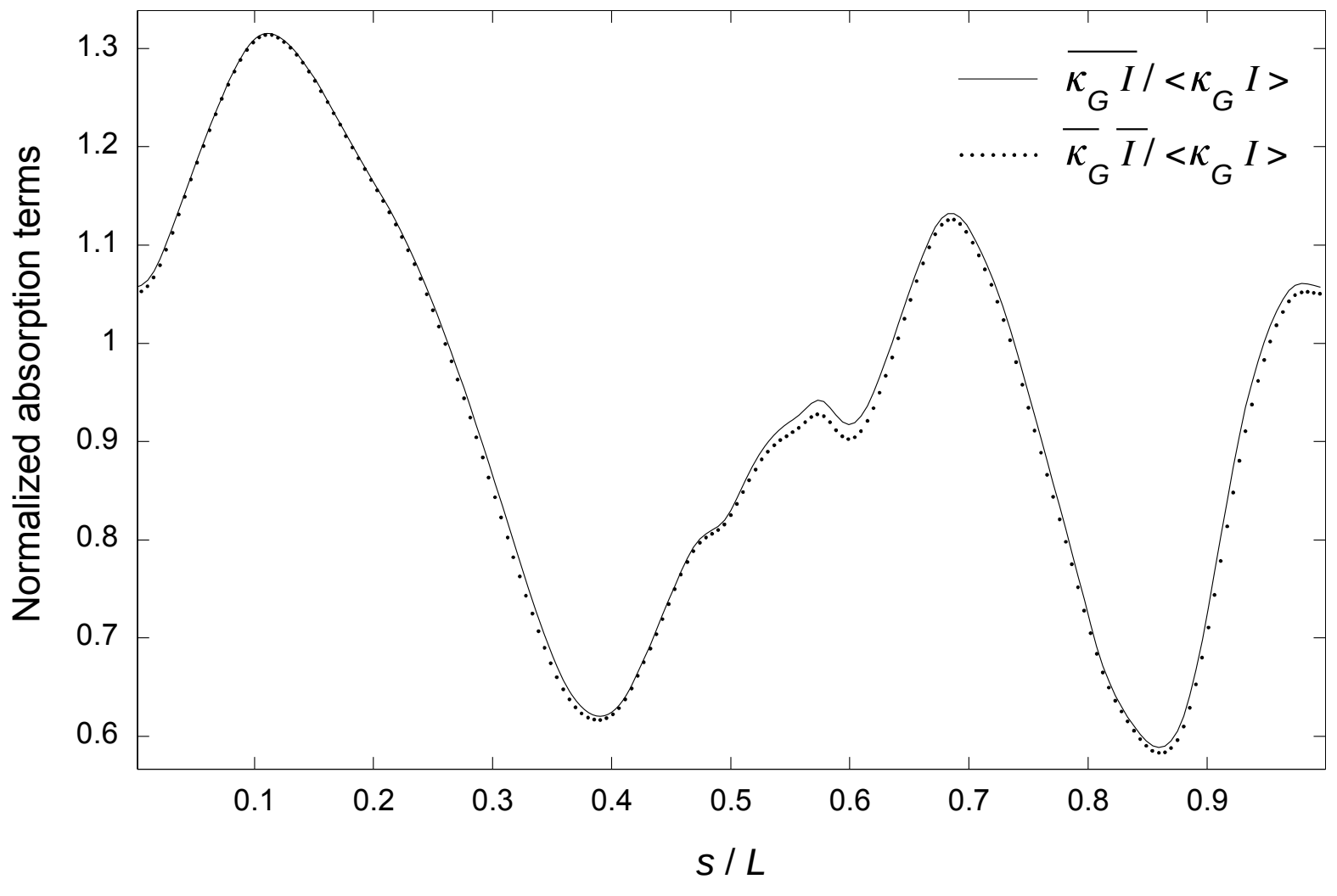

Figure 9. Absorption terms along a line of sight estimated by LES without subgrid-scale models compared with the filtered absorption terms determined from DNS (the filter size is Δ=16$\delta$).

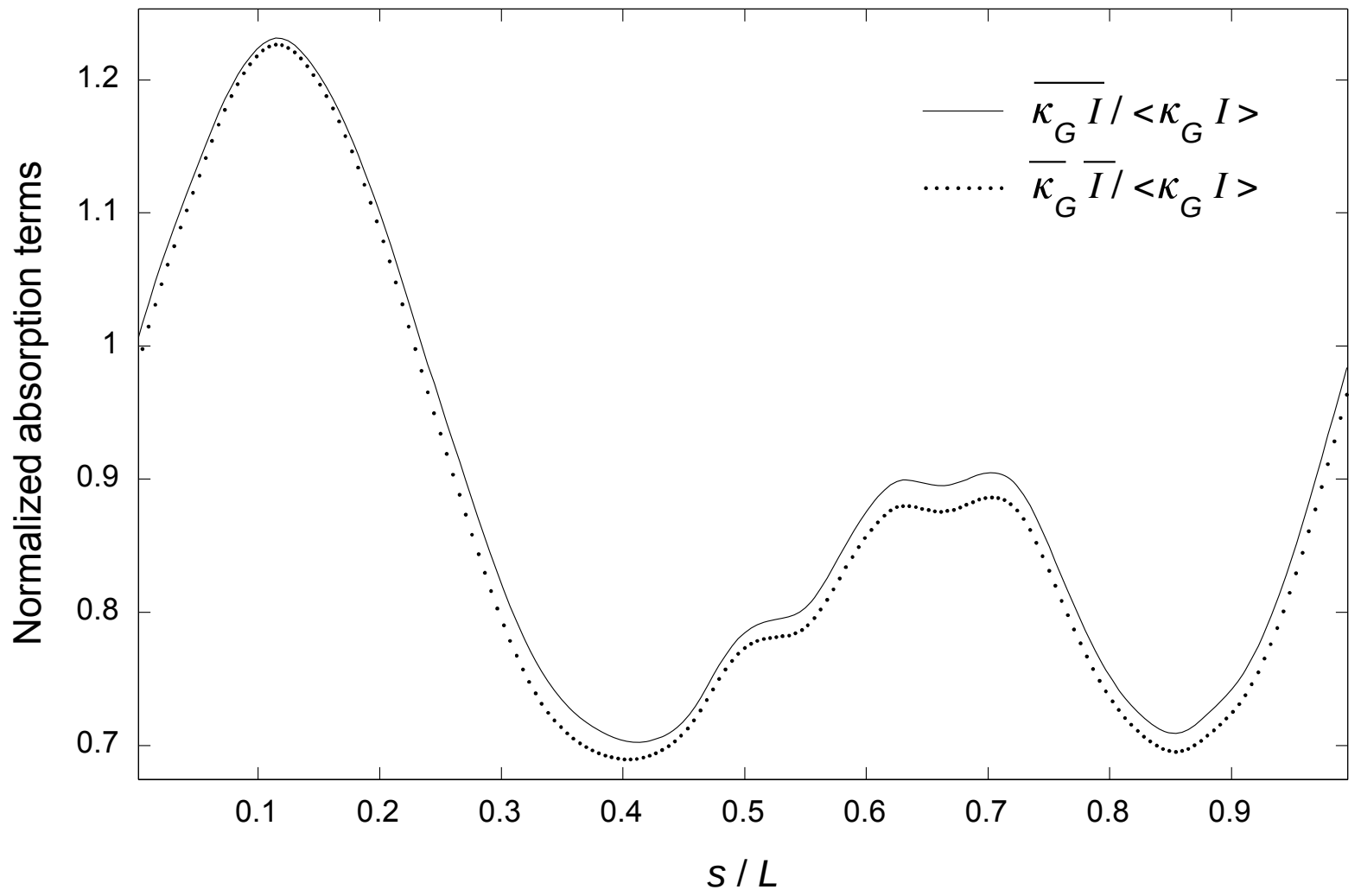

Figure 10. Absorption terms along a line of sight estimated by LES without subgrid-scale models compared with the filtered absorption terms determined from DNS (the filter size is $\Delta$=32$\delta$).

It was found that the need for subgrid-scale modelling was more important in the case of the emission terms of thermal radiation, because the absorption coefficient and the Planck function are local quantities, and consequently are more sensitive to local fluctuations than the absorption terms, which depend on the temperature and species concentration along the optical path. The turbulence intensity and the optical thickness of the filter are two important parameters that influence significantly the subgrid-scale correlations between the absorption coefficient and the radiation intensity or the blackbody intensity. Nevertheless, the assumption that consists in neglecting all subgrid-scale fluctuations has proved to be accurate in various cases, especially when the optical thickness of the filter is small (lower than 1), and when the turbulence intensity is lower than 20%. In other cases, this assumption should be used carefully, because local subgrid-scale fluctuation may not be negligible. In cases where the turbulence intensity is large as well as the optical thickness, the subgrid-scale fluctuations become important, and subgrid-scale models are needed.

This study has been based on an academic physical configuration, *i.e.* steady (forced) homogeneous isotropic turbulence and involve only one single time step, and therefore the extension of this analysis to real cases should be done very carefully. However, we emphasize that in turbulent reactive flows, the Reynolds number may be significantly higher than the one used in this study, and consequently the turbulence intensity may have a stronger influence. Moreover, in combustion systems, the optical thickness depends on the gas spectrum and may be higher than that evaluated in this study in various regions of the spectrum. Accordingly, in future works, we intend to propose and evaluate subgrid-scale models for turbulence-radiation interactions with the objective of including radiative transfer calculations in large eddy simulation of reactive flows.

## ACKNOWLEDGEMENTS

This work was developed within the framework of project POCI/EME/59879/2004, which is financially supported by FCT-Fundação para a Ciência e a Tecnologia, programme POCI 2010, partially funded by FEDER. M. Roger acknowledges the FCT-Fundação para a Ciência e a Tecnologia for the post-doc fellowship SFRH/BPD/34014/2006.